\documentclass[prl, superscriptaddress, twocolumn]{revtex4}
\usepackage{graphicx}
\usepackage[latin1]{inputenc}
\usepackage{amsfonts}
\usepackage{amsmath}
\usepackage[english]{babel}

\newcommand{\eps}{\varepsilon}
\newcommand{\Ham}{\mathcal{H}}

\begin{document}

\title{Atomic spin sensitive dissipation on magnetic surfaces}
\author{Franco Pellegrini}
\affiliation{International School for Advanced Studies (SISSA), Via Bonomea 265, I-34136 Trieste, Italy}
\affiliation{CNR-IOM Democritos National Simulation Center, Via Bonomea 265, I-34136 Trieste, Italy}
\author{Giuseppe E. Santoro}
\affiliation{International School for Advanced Studies (SISSA), Via Bonomea 265, I-34136 Trieste, Italy}
\affiliation{CNR-IOM Democritos National Simulation Center, Via Bonomea 265, I-34136 Trieste, Italy}
\affiliation{International Centre for Theoretical Physics (ICTP), P.O. Box 586, I-34014 Trieste, Italy}
\author{Erio Tosatti}
\affiliation{International School for Advanced Studies (SISSA), Via Bonomea 265, I-34136 Trieste, Italy}
\affiliation{CNR-IOM Democritos National Simulation Center, Via Bonomea 265, I-34136 Trieste, Italy}
\affiliation{International Centre for Theoretical Physics (ICTP), P.O. Box 586, I-34014 Trieste, Italy}

\date{\today}

\begin{abstract}
We identify the mechanism of energy dissipation relevant to spin-sensitive nanomechanics including the 
recently introduced magnetic exchange force microscopy, where oscillating magnetic tips approach surface atomic spins.
The tip-surface exchange couples spin and atom coordinates, leading to a spin-phonon problem with
Caldeira-Leggett type dissipation. In the overdamped regime, that can lead to a hysteretic flip of 
the local spin with a large spin-dependent dissipation, even down to the very low experimental tip oscillation 
frequencies, describing recent observations for Fe tips on NiO. 
A phase transition to an underdamped regime with dramatic
drop of magnetic tip dissipation should in principle be possible by tuning tip-surface distance.
\end{abstract}


\maketitle


In a recent intriguing magnetic exchange force microscopy experiment \cite{WiesRMP}, an
exquisite magnetic atomic force sensitivity was demonstrated for an atomically sharp Fe magnetic tip 
over the $(001)$ surface of antiferromagnetic NiO \cite{Wiesendanger}. 
Besides showing a different force for the two oppositely polarized surface Ni atoms ---
well explained by the Fe-Ni exchange available from electronic structure calculations\cite{Momida} ---
the results also show a surprisingly different mechanical dissipation, with a gigantic excess
of order 15-20 meV per cycle in the antiparallel Fe-Ni spin configuration, as compared
to the parallel one.  

There is no existing theory of spin-dependent tip dissipation that one could use to understand not 
just this result but magnetically and site sensitive dissipation phenomena in general. 
Here we propose to use the magnetic exchange force microscopy study as a starting point.   
We search for a mechanism that
i) can yield a magnetic dissipation of very large magnitude, similar to exchange energies, per cycle;
ii) is sensitive to the spin direction, and stronger for (nearly) antiparallel spin than a parallel one;
iii) works down to the lowest frequencies. 
Particularly puzzling is in fact the contrast between a large dissipation magnitude and the very low tip 
oscillation frequency ($\omega_{\rm tip}\sim 160$ kHz). At such a low frequency, one could expect a 
nearly adiabatic response, with very little mechanical energy transferred from the tip 
to some low-frequency excitations such as magnons, or perhaps phonons. 
Antiferromagnetic magnons, the first obvious choice, are immediately ruled out since,
owing to strong dipolar anisotropy, the antiferromagnetic spin-wave dispersion of NiO has a bulk 
gap $\Delta\sim1.5$ meV $\sim 0.36$ THz \cite{Hutchings},
and one at least as large at the surface \cite{DeWames,Mills,Schlecht}. 
As a result, the oscillatory perturbation exerted on the surface spin is completely adiabatic 
--- $\omega_{\rm tip} \ll \Delta$ by more than 6 orders of magnitude ---
and direct dissipation in the spin-wave channel vanishes. 
Other strictly magnetic dissipation mechanisms involving mesoscopic scale phenomena, such as domain 
wall motion~\cite{Liu1}, also appear unapplicable to the atomic scale tip-sample magnetic interaction. 
For example, a tip-induced magnetic domain with oscillating boundaries could 
be invoked to account for a low-frequency magnetic dissipation, but the formation of such local 
domains is energetically unlikely, given the localized nature of the tip perturbation: 
Excluding a role of tip stray fields, simple model estimates suggest that the spin deformation
near a perturbed surface spin should decay just a few atomic spacings away from the tip edge. 
We are left with acoustic phonons, certainly never gapped, both in bulk and at the surface.
Here we know however that acoustic dissipation of a localized surface oscillation vanishes in linear 
response theory as a high power of frequency\cite{Persson_Tosatti1999} -- the lattice can follow essentially 
adiabatically and harmonically a sufficiently slow and weak external perturbation. 
A large magnetic dissipation mechanism via phonons should therefore involve phenomena 
far from linear response. 
In this Letter we describe the mechanism which we believe is at work here, and
show that the nonlinear response is related to the attainment of a strong coupling
overdamped spin-phonon state very well known in other contexts, giving rise to 
a single-spin hysteresis. That also suggests that by tuning down the perturbation intensity, 
a phase transition could be crossed from the overdamped to the underdamped regime, 
with a loss of hysteresis and a dramatic drop of dissipation. Hopefully, the present approach may serve
as a prototype for nanoscale magnetic dissipation.

\begin{figure}[b]\centering
\includegraphics[width=.42\textwidth]{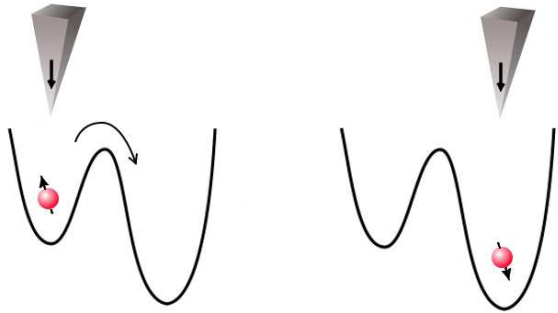}
\vspace{-10pt}
\caption{\label{SpinFlip}Effective potential felt by the spin under the effect of the bosonic bath 
for tip over spin-up and spin-down configuration.}
\end{figure}

Consider an oscillating Fe tip over a surface Ni spin $\vec{S}_i$. 
All neighboring spins remain essentially unperturbed, ``protected'' as they are by the spin gap $\Delta$.  
The potential felt by an $\uparrow$ Ni atom at a distance $z$ below the Fe tip 
differs from that felt by a $\downarrow$ Ni \cite{Momida}, 
and one can define a spin exchange potential 
$V^{\rm ex}(z)=V^{\downarrow\downarrow}(z)-V^{\uparrow\downarrow}(z)$ 
(assuming the Fe tip to be $\downarrow$ polarized).
$V^{\rm ex}$ can be estimated to yield an exchange force $f^{\rm ex}=-\partial V^{\rm ex}/\partial z$
of $\sim 0.3$ nN when the tip edge is closer than $3$ {\AA} to the surface Ni \cite{Momida}.
This force produces a small displacement $u_z(i)$ of the Ni atom from its equilibrium position and
will result in a potential of the form $-f^{\rm ex} S^z_i u_z(i)$ (neglecting an unimportant spin-independent term). 
In terms of phonon creation ($a^{\dagger}_{{\bf k}s}$) and annihilation ($a_{{\bf k}s}$) operators 
(${\bf k}s$ being wave vector and polarization of the phonon mode),
we thus obtain a coupling of the Ni-spin to the Ni acoustic phonons of the form 
$H_{\rm spin phonons}=\sigma^z_i \sum_{{\bf k}s} \lambda^{(i)}_{{\bf k}s} (a_{{\bf k}s} + a^{\dagger}_{-{\bf k}s})$,
where 
$\lambda^{(i)}_{{\bf k}s}=-f^{\rm ex} e^{i{\bf k}\cdot{\bf r}_i}\sqrt{\frac{\hbar}{8NM\omega_{{\bf k}s}}}e_z({\bf k}s)$;
$e_z({\bf k}s)$ being the eigenvector of the ${\bf k}s$ phonon mode.
The equilibrium physics of the spin is dictated by the small frequency behavior of the crucially important
spectral density \cite{Weiss,LeggRev}
$J(\omega)=\sum_{{\bf k}s} \delta(\omega-\omega_{{\bf k}s}) |\lambda_{{\bf k}s}|^2=(f^{\rm ex})^2 \frac{\hbar}{8MN}
\sum_{{\bf k}s} \delta(\omega-\omega_{{\bf k}s}) |e_z({\bf k}s)|^2/ \omega_{{\bf k}s}$.
From the standard Debye form for the low-energy acoustic phonons in three-dimensions, we 
find that the small-$\omega$ limit of $J(\omega)$ is precisely Ohmic $J(\omega)=\hbar^2\alpha\omega + \cdots$ 
with $\alpha=(f^{\rm ex})^2 \frac{3\hbar^2}{8Mk^3T_D^3}$, where $T_D$ is the Debye temperature. 
An estimate, with $f^{\rm ex}\sim 0.3$ nN, gives a value of $\alpha$ close to $1$, which can be easily 
made $>1$ by a slightly larger $f^{\rm ex}$ or by a better account of the (softer) surface phonon modes.
The natural Ohmic behavior of this problem is a first important result, since that is by far the
most interesting case, studied for decades \cite{Weiss,LeggRev}, and it has been previously shown to arise in
tip-surface interactions \cite{LouisSethna}. 
Note that the Ohmic coupling $\alpha$ depends on the square of the exchange force $f^{\rm ex}$ and is
therefore dependent on the tip-atom distance $z$. We are thus led to the physics of a single spin
--- the surface Ni over which the tip is oscillating --- in its prototypical form, that of 
a driven Caldeira-Leggett (or spin-boson) Ohmic model.
The model is known to possess two regimes, one underdamped and one overdamped, separated by a phase
transition. 
In the underdamped regime the spin motion is relatively unaffected by the bath, and dissipation is small. 
In the overdamped regime --- attained at $\alpha>1$ \cite{LeggRev} ---
the spin is effectively ``trapped'' by the bath as schematically portrayed in Fig.~\ref{SpinFlip}. 

%
The tip-induced spin-flip processes 
involve in the overdamped case a new time scale $\gamma^{-1}$ that can be much longer than the 
external driving period $2\pi \omega_{\rm tip}^{-1}$. Once $\gamma\ll \omega_{\rm tip}$, magnetic dissipation 
will arise from a sort of single-spin hysteresis, similar to effects known in bistable models \cite{Thorw}. 
The overdamped model predicts three crucial results regarding spin-dependent dissipation.
First, dissipation is quantitatively large, because the order of magnitude of the hysteresis loop area 
is generally set by the tip-surface magnetic exchange, itself a large energy scale $\sim 50$ meV \cite{Momida}. 
Second, dissipation will be strong when the magnetic tip is over a surface atom with antiparallel spin 
(left-hand side of Fig.~\ref{SpinFlip}), and negligible over one with parallel spin (right-hand side of Fig.~\ref{SpinFlip}), 
because no tip-induced spin-flip is provoked in the latter. 
Third, hysteretic dissipation should depend relatively weakly on $\omega_{\rm tip}$, and disappear only when the 
tip frequency is lowered below some very low frequency $\sim \gamma$, 
itself temperature-dependent.
 
To describe the action of the tip on a given spin, we consider the driven spin-boson model ($\hbar=1$):
\begin{eqnarray} \label{Ham}
\Ham &=& \frac{\eps_0}{2}\sigma_z - \left[\frac{\eps(t)}{2}\sigma_z+\frac{\Delta(t)}{2} \sigma_x\right] 
        -\frac{\hat{X}}{2} \sigma_z \nonumber \\
&& + \sum_{\nu} \omega_{\nu} \left(b_{\nu}^\dag b_{\nu}+\frac{1}{2}\right) \;,
\end{eqnarray}
where $\sigma_z$ and $\sigma_x$ are the Pauli matrices, $b_{\nu}^\dag$ and $b_{\nu}$ are creation and 
annihilation operators for the mode $\nu$ of the phononic bath of frequency $\omega_{\nu}$, 
$\hat{X} = \sum_{\nu} \lambda_{\nu} \left(b_{\nu}+b_{\nu}^\dag\right)$ is the bath operator
to which $\sigma_z$ is coupled, $\lambda_{\nu}$ being the previously introduced couplings, 
such that $J(\omega)=\sum_{\nu} \delta(\omega-\omega_{\nu}) |\lambda_{\nu}|^2=\alpha \omega + \cdots$.
(The slight time dependence of $\alpha$ during the tip oscillation is neglected.)
In the NiO surface, $\eps_0$ represents nearest-neighbor exchange (positive and large) 
plus all sources of magnetic anisotropy (including dipolar contributions), and
$z$ is the unperturbed direction of the surface atom spin due to dipolar anisotropy 
(a $\left\langle 211\right\rangle$ direction).
The external magnetic tip acts in the $xz$ plane \cite{Wiesendanger} at an angle $\theta$ off the $z$ axis,
\begin{equation} \label{ExtP}
\eps(t)=h\cos\theta\sin^2(\omega_{\rm tip} t), \; \Delta(t)=h\sin\theta\sin^2(\omega_{\rm tip} t) \;.
\end{equation}
We will use $\theta$ and $\theta+\pi$ to describe Ni spins of opposite direction.
The last term describes the free phonons. 
To avoid problems with divergences, we have as usual a high-frequency cutoff $\omega_c$ for the spectral density, 
$J(\omega)=\alpha\omega e^{-\omega/\omega_c}$, with $\omega_c\gg h,\,k_BT$. 
The details of the solution of this model, summarized below for the reader's convenience, are standard. 
The time evolution of the system is described by a standard real-time path-integral approach \cite{Weiss}, taking care of the bath degrees of freedom through the influence-functional method \cite{FeyVern} and applying the so-called noninteracting-blip approximation, valid in the $\alpha>1$ overdamped case for the observables of our interest \cite{caldeira_leggett,LeggRev}. Actually, the same overdamped behavior could be achieved in the $\alpha<1$ regime, but only for certain values of perturbation and temperature.
In terms of the free correlation function of the $X$ operator $g(\tau)=S(\tau)+iR(\tau)$, 
with $S(\tau) = \int_0^\infty \! d\omega \; 
\frac{J(\omega)}{\omega^2}(1-\cos \omega\tau)\coth\left(\frac{\beta\omega}{2}\right)$ 
and $R(\tau) = \int_0^\infty \! d\omega \; \frac{J(\omega)}{\omega^2}\sin \omega\tau$ 
(where $\beta=1/k_BT$), we can compute the quantities
$F_0(t) =\Delta^2(t)\int_0^\infty \! d\tau \; 
                   e^{-S(\tau)} \sin{[R(\tau)]} \sin{\left[ \varepsilon(t)\tau\right]} $ and
$G_0(t) =\Delta^2(t)\int_0^\infty \! d\tau \; 
                   e^{-S(\tau)} \cos{[R(\tau)]} \cos{\left[ \varepsilon(t)\tau\right]} $.
From these, as shown in \cite{GrifPRE48}, in the overdamped regime $\alpha>1$ and with a low-frequency driving, 
the $z$ component of the spin obeys a simple rate equation
\begin{equation} \label{SzRE}
\frac{d}{dt}\left\langle \sigma_z(t)\right\rangle=-G_0(t)\left\langle \sigma_z(t)\right\rangle+F_0(t) \;,
\end{equation}
that can be easily integrated.
Applying the same procedure to $\left\langle \sigma_x(t)\right\rangle$, we get similarly 
\begin{equation} \label{SxRE}
\frac{d}{dt}\left\langle \sigma_x(t)\right\rangle = -G_0(t)\left\langle \sigma_x(t)\right\rangle
                          + \frac{F_0(t) \tilde{G}_0(t)}{\Delta(t)} \;,
\end{equation}
where $\tilde{G}_0(t)$ is defined as $G_0(t)$, but with $\sin{\left[ \varepsilon(t)\tau\right]}$ 
in place of $\cos{\left[ \varepsilon(t)\tau\right]}$. 

To uncover the new time scale, consider, e.g., the abrupt switching on of a perturbation at $t=0$, 
with a constant value $\epsilon(t)=\bar{\varepsilon}-\eps_0$ and $\Delta(t)=\Delta$ thereafter. 
In this case Eq.~\eqref{SzRE} describes an exponential relaxation towards the equilibrium value 
$\left\langle \sigma_z(\infty)\right\rangle=-F_0/G_0= -\tanh(\frac{1}{2}\beta\bar{\varepsilon})$ \cite{LeggRev} 
with a decay rate $\gamma$ given, for low temperatures, in terms of the $\Gamma$ function: 
$\gamma=\left(\pi\,\Delta^2\,\bar{\eps}^{2\alpha-1}/2\,\Gamma(2\alpha)\,\omega_c^{2\alpha}\right)
\left[1+\frac{\pi^2\alpha(2\alpha-1)(2\alpha-2)}{3(\beta\bar{\eps})^2}\right]$. 
In the overdamped regime the relevant time scale $\gamma^{-1}$ can take 
large values, mainly due to the large cutoff frequency $\omega_c$ being raised to a large exponent.
Even under a comparably slow external perturbation the system can be out 
of its instantaneous equilibrium; and that is the origin of the hysteretic behavior.

\begin{figure}[!t]\centering
\includegraphics[width=.47\textwidth]{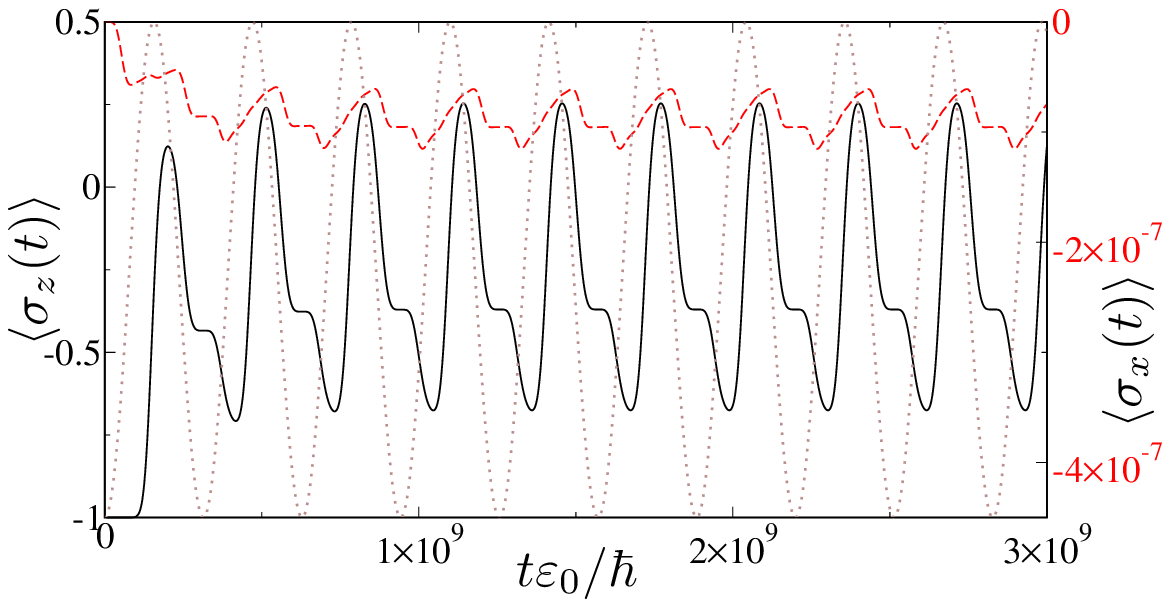}
\vspace{-10pt}
\caption{\label{Data1}(Color online) 
  Time evolution of $\left\langle \sigma_z\right\rangle$ (black full line, left-hand axis) 
  and $\left\langle \sigma_x\right\rangle $ (dashed red line, right-hand axis) for $\eps_0=1$, $\omega_c=20\eps_0$, 
  $\alpha=2.1$, $h=1.5\eps_0$, $\theta=0.6$, $\omega_{\rm tip}=10^{-8}\eps_0$, $\beta\eps_0=20$.
  The dotted line shows the shape of the external perturbation.}
\vspace{-7pt}
\end{figure}

\begin{figure}[!b]\centering
\includegraphics[width=.42\textwidth]{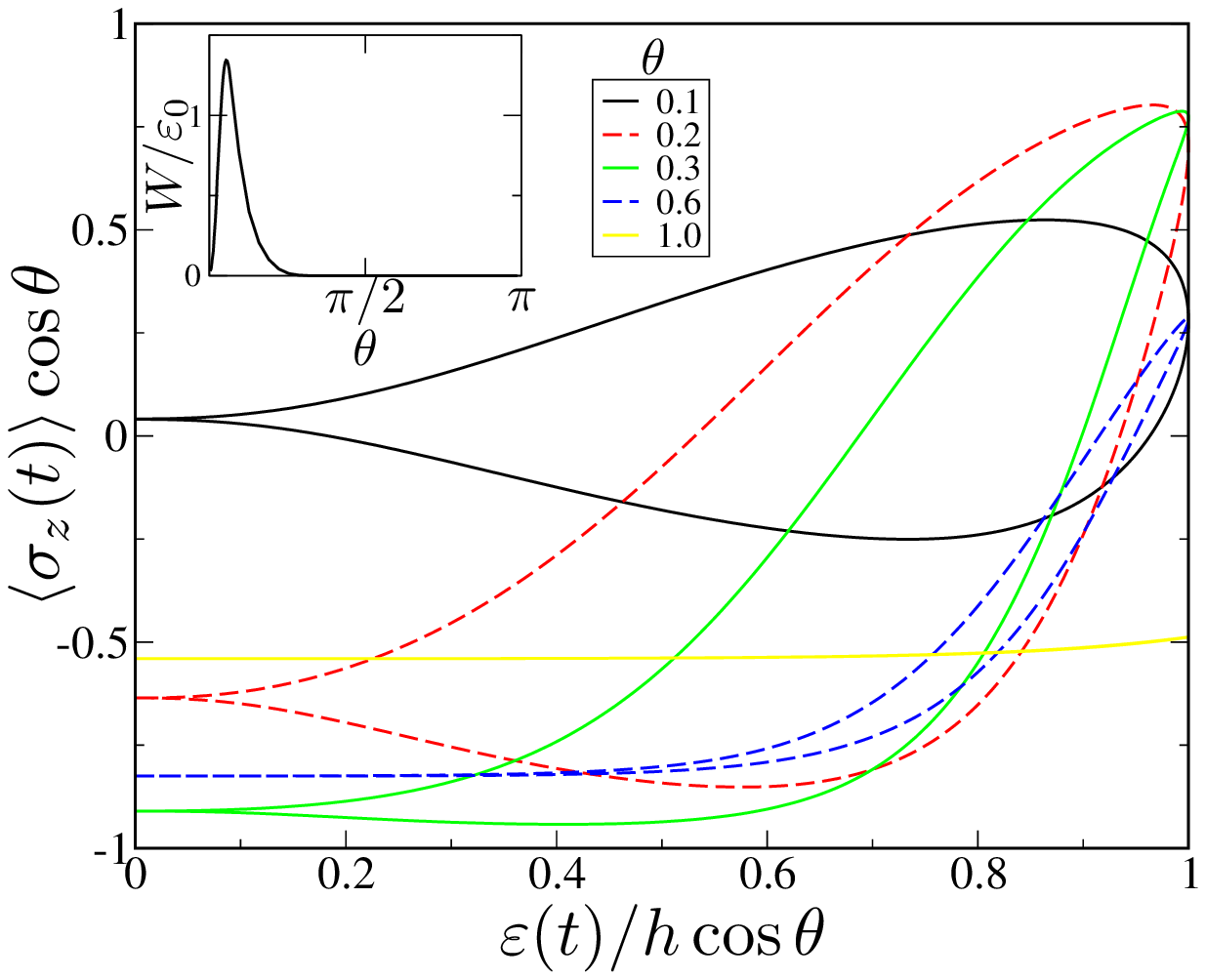}
\vspace{-7pt}
\caption{\label{Hyst1}(Color online) 
 Hysteresis cycle for the $z$ component of the external perturbation for different angles $\theta$ (see legend) 
 for $\eps_0=1$, $\omega_c=100\eps_0$, $\alpha=2.1$, $h=1.3\eps_0$, $\omega_{\rm tip}=10^{-12}\eps_0$. 
 Inset: Angular dependence of the hysteresis area $W$.}
\end{figure}

Consider now the two-level system of Eq.~\eqref{Ham} under the external perturbation in \eqref{ExtP}. 
Figure \ref{Data1} shows the time evolution of $\left\langle \sigma_z\right\rangle$ and 
$\left\langle \sigma_x\right\rangle$ for a system in the initial state $\left\langle \sigma_z\right\rangle=-1$, 
as obtained by numerical integration of Eqs.~\eqref{SzRE} and \eqref{SxRE} using an adaptive 
Runge-Kutta algorithm \cite{NumRec}. 
The system can clearly respond to the perturbation in its own time scale, 
yielding two different states when the perturbation is increasing and decreasing, a 
clear hysteretic behavior. The values of $\left\langle\sigma_x(t)\right\rangle$ are orders of magnitude
smaller than those of $\left\langle\sigma_z(t)\right\rangle$ due to the large $\omega_c$. 
[This result is related to the
universal behavior of $\left\langle\sigma_z(t)\right\rangle$ with respect to $\omega_c$ as opposed
to the nonuniversality of $\left\langle\sigma_x(t)\right\rangle$ \cite{Guinea}.]
The dissipated energy per cycle is 
\begin{equation} 
W = \int_0^{2\pi/\omega} \!\!\! dt \; \left[\left\langle \sigma_z(t)\right\rangle\frac{d\eps(t)}{dt} 
                               + \left\langle\sigma_x(t)\right\rangle\frac{d\Delta(t)}{dt}\right] \;,
\end{equation}
which is the area of the hysteresis cycle 
in a (state-perturbation) $\langle \sigma_z(t)\rangle$-$\epsilon(t)$ diagram.
Figure \ref{Hyst1} shows the hysteresis cycle for the $z$ component of perturbations 
with different angles $\theta$, together with the angular dependence of the hysteresis area (inset). 
When the tip and atom spins are (even roughly) opposite, the action of exchange to overturn the spin
leads to a hysteresis loop and a large dissipation; when they are nearly parallel, the loop collapses and 
correspondingly the magnetic tip dissipation drops. (Data for the $x$ component, not shown, are negligible.)

%
%

\begin{figure}[!t]\centering
\includegraphics[width=.47\textwidth]{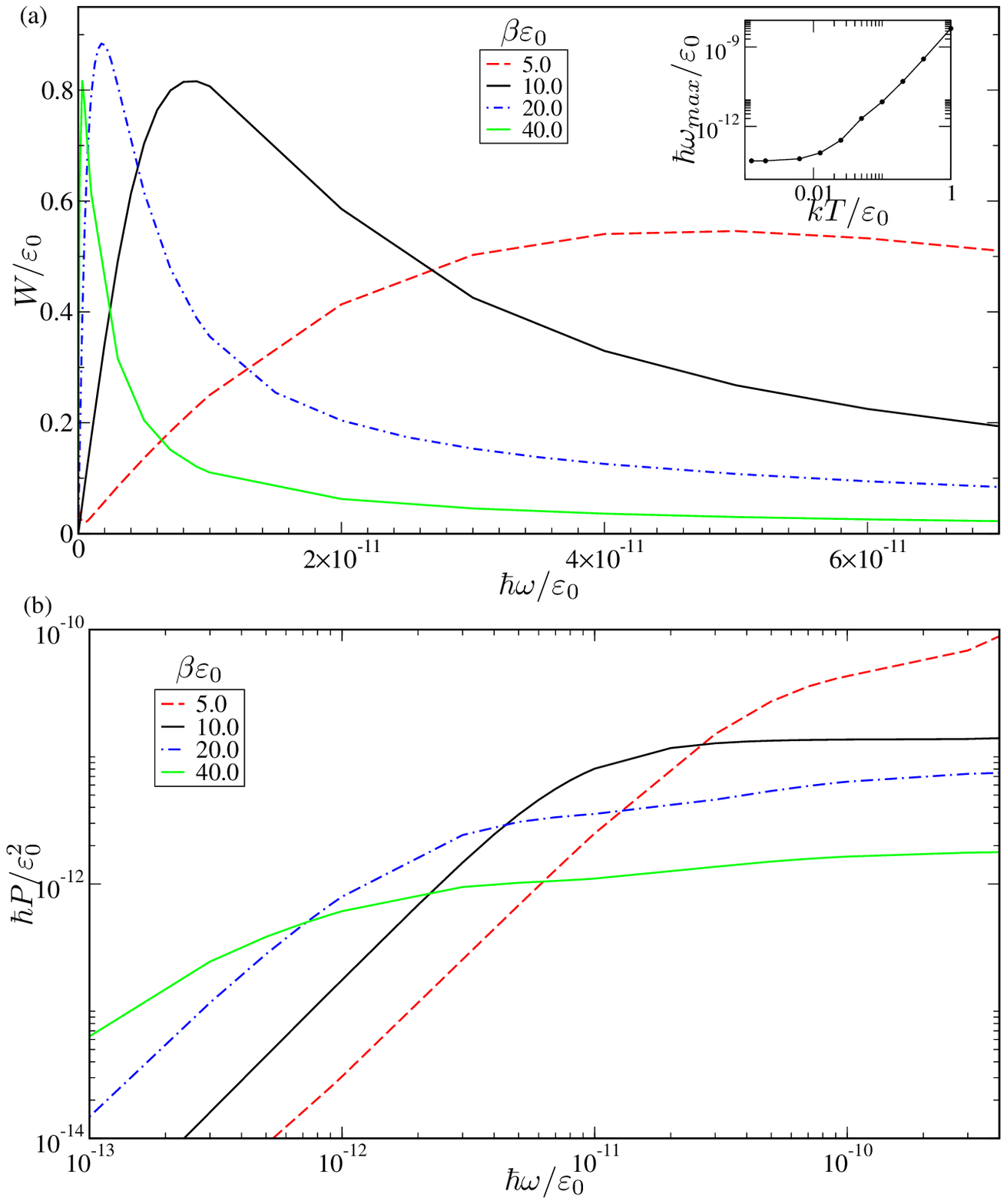}
\caption{\label{Diss}(Color online) 
  (a) Area of the hysteresis loop as a function of frequency for various values of $\beta$ (see legend). 
      Inset: Frequency of maximum area of the hysteresis loop as a function of the temperature $T$. 
  (b) Dissipated energy as a function of frequency for various values of $\beta$ (see legend).}
\end{figure}

We may finally address the frequency and temperature dependence of the total magnetic dissipation. 
Loop areas (in steady state) for different temperatures as a function of frequency are shown in Fig.~\ref{Diss}(a). 
There clearly is an optimal frequency attaining maximal area. At excessive tip frequencies the spin remains effectively frozen in its trapped state; at very low frequencies the spin has plenty of time to relax and follow adiabatically the equilibrium value demanded by the tip: in both cases the loop area collapses.
The inset of Fig.~\ref{Diss}(a) shows how the optimal frequency increases with increasing temperature,
reflecting the $T$ dependence of $\gamma$ shown in the forcing-free case. 
These results are consistent with what is known in the context of quantum stochastic resonance \cite{Gammait}.
Figure \ref{Diss}(b) contains the same data in the form of $P=\omega W$, the dissipated power. 
At low frequencies $P$ increases as a power law (roughly $\omega^2$), 
eventually reaching a plateau where the dissipation levels off over 
a wide frequency range. 
Other mechanisms will of course play a role at higher frequencies, but hysteretic dissipation 
is the only relevant nonlinear one that survives down to experimentally relevant low frequencies.  

The magnetic dissipation per cycle produced by the mechanism identified 
satisfies all desired requisites, since it is 
(i) large, and of the same order of magnitude of the antiferromagnetic exchange $J\approx$ 15-20 meV, 
(ii) vastly different for ``up'' and ``down'' Ni spins 
(assuming $\theta$ = 35°, we get 
between the two a factor of $3\times 10^{-4}$), 
and (iii) effective down to very low frequencies, $\sim \gamma$.
Coming to the Fe-NiO data, we can now attribute the experimental dissipation 
of about $35$ meV of the $\downarrow$-polarized tip oscillating over a $\downarrow$ Ni spin to nonmagnetic mechanisms, 
that of about 50 meV over an $\uparrow$ Ni in terms of the same mechanism {\em plus} a hysteretic magnetic 
dissipation $W \approx$ 15 meV, implying that the tip-surface coupling resulted in $\alpha>1$.

The strong dependence of that coupling on $f_{\rm ex}$ allows in principle for a reduction of $\alpha$
and a phase transition from overdamped to underdamped. In that case we would expect the faster evolution 
timescale to suppress the hysteretic behavior, dramatically reducing the magnetic dissipation. 

In conclusion, our main novelties are that in surface magnetic tip dissipation problems, energy dissipation
should be mostly mechanical and non-spin-wave, since spin waves are generally gapped by anisotropy; 
that spin-dependent coupling to surface atomic motion and to phonons can lead to a 
sort of single-site magnetic hysteresis; and finally that due to hysteresis the magnetic tip dissipation per cycle 
can be as large as intrasurface exchange coupling, as is seen experimentally. 
Further experimental possibilities will be to test the frequency, temperature and $\theta$ angle dependencies.
Last but not least, the modification of the coupling parameter with the tip-sample interaction force 
should in principle cause a phase transition from an overdamped to an underdamped regime, with
a strong suppression of dissipation at large distances and weaker couplings. We believe that
these concepts should be of more general impact beyond the simple case treated here. 
\begin{acknowledgments}
This work is supported by the CNR/FANAS project AFRI under Eurocores ESF, and by a PRIN-COFIN contract of the Italian
University and Research Ministry. Informative discussions with Professor R. Wiesendanger are gratefully acknowledged.
\end{acknowledgments}
%



\end{document}